\documentclass[prl,twocolumn,aps,floatfix,preprintnumbers,groupedaddress,nofootinbib]{revtex4-1}
\pdfoutput=1

\newcommand{\slashed}[1]{\not\!#1}

\usepackage{amsmath}
\usepackage{amssymb, bm}
\usepackage{graphicx}
\usepackage{float}
\usepackage{braket}
\usepackage{hyperref}
\usepackage{pbox}
\usepackage{tabularx}
\usepackage{enumitem}
\usepackage{tikz}
\usepackage{tkz-euclide}
\usetikzlibrary{intersections}
\usetikzlibrary{trees}
\usetikzlibrary{decorations.pathmorphing}
\usetikzlibrary{decorations.markings}
\usetikzlibrary{decorations, decorations.markings, decorations.pathmorphing, arrows, graphs, shapes.geometric, snakes}
\usetikzlibrary{arrows}
\usetkzobj{all}
\def\radius{1.mm} 
\tikzset{
electron/.style={draw=violet, postaction={decorate},
        decoration={markings,mark=at position .55 with {\arrow[draw=violet]{>}}}},
quark/.style={draw=black, postaction={decorate},
        decoration={markings,mark=at position .55 with {\arrow[draw=black]{>}}}},
antiquark/.style={draw=black, postaction={decorate},
        decoration={markings,mark=at position .55 with {\arrow[draw=black]{<}}}},
quarku/.style={draw=black, postaction={decorate},
        decoration={markings,mark=at position .75 with {\arrow[draw=black]{>}}}},
antiquarku/.style={draw=black, postaction={decorate},
        decoration={markings,mark=at position .75 with {\arrow[draw=black]{<}}}},
        gluon/.style={decorate, draw=or,
        decoration={coil,amplitude=2pt, segment length=3pt}},
  }
\definecolor{greeen}{rgb}{0.03,0.84,0.13}
\definecolor{test}{rgb}{0.03,0.74,0.33}
\definecolor{viol}{rgb}{0.44,0,0.94}
\definecolor{or}{rgb}{0.95,0.65,0}

\newcommand{\bmat}{\begin{pmatrix}}
\newcommand{\emat}{\end{pmatrix}}
\def\beq{\begin{equation}}
\def\eeq{\end{equation}}
\def\beeq{\begin{eqnarray}}
\def\eeeq{\end{eqnarray}}
\def\bea{\begin{align}}
\def\eea{\end{align}}

\def\gp2{g^{\prime 2}}

\usepackage{color}
\begin{document}   
\title{$R$-parity Violating Supersymmetry at IceCube} 
\author{P. S. Bhupal Dev$^1$, Dilip Kumar Ghosh$^2$, Werner Rodejohann$^1$} 
\affiliation{$^1$ Max-Planck-Institut f\"{u}r Kernphysik, Saupfercheckweg 1, D-69117 Heidelberg, Germany}
\affiliation{$^2$ Department of Theoretical Physics, Indian Association for the Cultivation of Science,
2A \& 2B, Raja S.C. Mullick Road, Kolkata 700032, India}
\begin{abstract}
The presence of $R$-parity violating (RPV) supersymmetric interactions 
involving high-energy neutrinos can lead to resonant production of TeV-scale 
squarks inside large-volume neutrino detectors. Using the ultra-high energy 
neutrino events observed recently at the IceCube, with the fact that for a 
given power-law flux of astrophysical neutrinos, there is no statistically 
significant deviation in the current data from the Standard Model 
expectations, we derive robust upper limits on the RPV couplings as a 
function of the resonantly-produced squark mass, independent of the 
other unknown model parameters, as long as the squarks decay dominantly to 2-body final states involving leptons and quarks through the RPV couplings. With more statistics, we expect these limits to be comparable/complementary to the existing limits from direct collider searches 
and other low-energy processes. 
\end{abstract}
\maketitle
\section{Introduction}
The detection of ultra-high energy (UHE) neutrinos at the 
IceCube~\cite{Aartsen:2013bka, Aartsen:2015zva} has opened a new window on 
the Universe. In the 4-year dataset, 54 UHE neutrino events have been observed,
which constitute a $6.5\sigma$ excess over the expected atmospheric 
background~\cite{Aartsen:2015zva}. It is imperative for both 
astrophysics and particle physics communities to understand all possible 
aspects of these UHE neutrino events reported by the IceCube Collaboration. 
From the astrophysics side, one needs to identify the possible extraterrestrial source(s)~\cite{Murase:2014tsa}, and the underlying spectral shape~\cite{Adrian-Martinez:2015ver} and flavor composition~\cite{Aartsen:2015knd}, 
of the neutrino flux. From a particle physics point of view, one can use this as a unique opportunity to test the Standard Model (SM) at energy scales that are otherwise 
not achievable on Earth~\cite{Anchordoqui:2013dnh}. So far, no statistically significant deviations from the SM prediction have been found in the IceCube data~\cite{Aartsen:2015zva, Kistler:2013my, Laha:2013lka, Chen:2013dza, Vincent:2016nut}, although many new physics scenarios have been envisaged (see Ref.~\cite{Anchordoqui:2013dnh} for an overview) to explain some peculiar features. With more statistics, if the data remains consistent with the SM predictions, one can put useful,  complementary constraints on various new physics scenarios. Anticipating this, we examine the current and future prospects of a well-motivated new physics scenario, namely,  $R$-parity violating (RPV) supersymmetry (SUSY)~\cite{Barbier:2004ez}, at the IceCube and beyond.   

The SUSY extension of the SM has many attractive features to qualify arguably as the best motivated candidate for the 
new physics~\cite{Haber:1984rc}. However, the lack of evidence for superpartners in the LHC data so far~\cite{atlas-susy, cms-susy} has forced the simplest SUSY scenarios toward regions of parameter space unnatural for the Higgs sector~\cite{Papucci:2011wy, Hall:2011aa, Evans:2013jna, Drees:2015aeo}. A simple way to preserve the Higgs naturalness by evading the current experimental constraints is by allowing RPV in the production and decays of superpartners~\cite{Brust:2011tb}. Apart from significantly lowering the collider bounds on the SUSY spectrum~\cite{Hall:1983id, Carpenter:2006hs, Allanach:2012vj, Asano:2012gj, Graham:2014vya}, RPV SUSY implies the violation of baryon and/or lepton numbers, which has important phenomenological consequences~\cite{Barbier:2004ez}. For instance, one can automatically generate non-zero neutrino masses and mixing~\cite{Hall:1983id, Dawson:1985vr, Ellis:1984gi, Joshipura:1994ib, Hempfling:1995wj, Roy:1996bua, Mukhopadhyaya:1998xj, 
 Bhattacharyya:1999tv, Davidson:2000ne, Diaz:2003as, Dreiner:2011ft} at either tree or one-loop level without introducing any extra particles beyond the Minimal Supersymmetric SM (MSSM) field content. Similarly, the presence of $\Delta L\neq 0$ RPV vertices can lead to observable neutrinoless double beta decay ($0\nu\beta\beta$)~\cite{Mohapatra:1986su, Vergados:1986td, Hirsch:1995zi, Babu:1995vh, Pas:1998nn, Faessler:2007nz, Allanach:2009xx, Kom:2011nc, Rodejohann:2011mu}, as well as successful baryogenesis~\cite{Cline:1990bw, Masiero:1992bv, Sarkar:1996sn, Dolgov:2006ay, Cui:2012jh, Arcadi:2015ffa, Sorbello:2013xwa, Adhikari:2016yyu}. Moreover, RPV scenarios also provide a viable explanation for a number of recent $2\sigma$ to $4\sigma$ anomalies, e.g. muon anomalous magnetic moment~\cite{Kim:2001se, Bhattacharyya:2009hb, Hundi:2011si, Chakraborty:2015bsk}, in semileptonic $B$-meson~\cite{Alok:2009wk, Deshpande:2012rr, Biswas:2014gga, Huang:2015vpt, Bauer:2015knc, Zhu:2016xdg} and lepton-flavor-violating Higgs~\cite{Arhrib:2012ax} decays, CMS $eejj$ and $e\nu jj$ excesses~\cite{Biswas:2014gga, Allanach:2014lca}, ATLAS diboson excess~\cite{Allanach:2015blv, Bhattacherjee:2015svr}, and the LHC diphoton excess~\cite{Ding:2015rxx, Allanach:2015ixl}. If any of these anomalies persist and become statistically more significant, one should consider RPV SUSY as a strong contender for the underlying new physics. In any case, it is of paramount importance to find complementary ways at as many different energy scales as possible to test this scenario. 
 
The most general RPV superpotential in the MSSM is 
\begin{align}
W_{\slashed{R}} \ = \ & \mu_i L_i H_u + \frac{1}{2}\lambda_{ijk}L_iL_jE_k^c+\lambda'_{ijk} L_i Q_j D_k^c \nonumber \\
& \qquad 
+\frac{1}{2}\lambda''_{ijk}U_i^c D_j^c D_k^c \, ,
\label{supRPV}
\end{align}
where $L_i\ni (\nu_i, e_i)_L$ and $Q_i \ni (u_i, d_i)_L$ are the $SU(2)_L$-doublet and $U_i^c, D_i^c, E_i^c$ are the $SU(2)_L$-singlet chiral superfields, respectively (with $i,j,k=1,2,3$ being the generation indices) and $H_u$ is the up-type Higgs superfield. Here we have suppressed all gauge indices for brevity. $SU(2)_L$ and $SU(3)_c$ gauge invariance enforce antisymmetry of the $\lambda_{ijk}$- and $\lambda''_{ijk}$-couplings with respect to their first and last two indices, respectively. Since we are interested in the UHE neutrino interactions with nucleons, we will only focus on the $\lambda'_{ijk}$-couplings. Any of these 27 new dimensionless complex parameters can lead to resonant production of TeV-scale squarks at IceCube energies~\cite{Carena:1998gd}, thereby making a potentially significant contribution to the UHE neutrino events. Note that even without SUSY, similar resonance features in neutrino-nucleon interactions can also occur in models with TeV-scale leptoquarks~\cite{Berezinsky:1985yw, Robinett:1987ym, Anchordoqui:2006wc, Alikhanov:2013fda, Barger:2013pla, Dutta:2015dka, Dey:2015eaa}.\footnote{However, there are subtle differences between scalar leptoquark and RPV SUSY models, e.g. due to the presence of additional decay channels and chiral mixing between squarks in the RPV case. Moreover, the recent papers analyzing the IceCube data in the context of leptoquark models have not taken into account the LHC and low-energy constraints.} The $\lambda$-couplings in Eq.~\eqref{supRPV} can give rise to resonant production of selectrons from neutrino interactions with electrons, reminiscent of the Glashow resonance~\cite{Glashow:1960zz} in the SM; however, for TeV-scale selectrons, their contribution to the total number of IceCube events is negligible in the energy range considered here and we will comment on this possibility later. Note that with non-zero $\lambda'$-couplings, we need to explicitly forbid the $\lambda''$-terms, e.g.~by imposing baryon triality~\cite{Ibanez:1991hv}, to avoid rapid proton decay~\cite{Smirnov:1996bg, Nath:2006ut}. We also ignore the bilinear terms in Eq.~\eqref{supRPV}, since they do not give rise to the resonance feature exploited here.\footnote{However, they  could lead to other distinct signatures (e.g. triple bang) relevant for future multi-km$^3$ neutrino telescopes~\cite{Hirsch:2007kx}.}

Using the fact that for a given power-law astrophysical neutrino flux, there is no statistically significant resonance-like feature in the current IceCube high-energy starting event (HESE) data, we derive robust upper limits on the RPV couplings $|\lambda'_{ijk}|$ as a function of the resonantly-produced down-type squark mass $m_{\tilde{d}_k}$, independent of the other SUSY parameters, provided the squarks decay dominantly through their RPV couplings. With the currently available low statistics of the IceCube HESE data, our bounds turn out to be weaker than the existing indirect constraints~\cite{Barbier:2004ez, Allanach:1999ic, Kao:2009fg, Dreiner:2012mx} from precision measurements in various low-energy processes. However, with more data pouring in from IceCube, and with the possibility of a second km$^3$ detector, such as KM3Net~\cite{Adrian-Martinez:2016fdl}, and even a multi-km$^3$ extension, such as IceCube-Gen2~\cite{Aartsen:2014njl}, our projected future limits could be comparable to the existing ones and  complementary to the direct probes of the $\lambda'$-type RPV SUSY at the LHC~\cite{Choudhury:2002av, Redelbach:2015meu}, as well as other indirect searches at future low-energy neutrino experiments~\cite{Barranco:2007tz}. This should provide yet another science motivation for the next-generation neutrino telescopes, or at least should allow for an independent test of a possible finding in the LHC or other experiments. 

\section{Neutrino-Nucleon Interactions} 
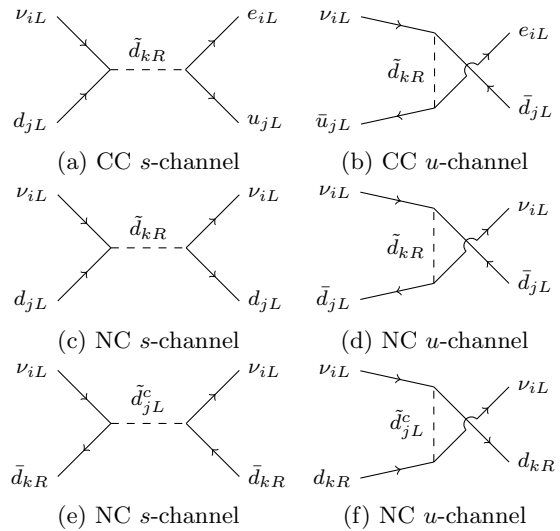
\begin{figure}[t]
  \centering
\begin{tabular}{cc}
  \noindent
  \begin{tikzpicture}[]
   \draw[quark] (-1.207,0.707)node[left]{{\footnotesize$\nu_{iL}$}} -- (-0.5,0);
  \draw[quark] (-1.207,-0.707)node[left]{{\footnotesize$d_{jL}$}} -- (-0.5,0);
  \draw[dashed](-0.5,0)--(0,0)node[above]{{\footnotesize$\tilde{d}_{kR}$}} -- (0.5,0);
  \draw[quark](0.5,0)--(1.207,-0.707)node[right]{{\footnotesize$u_{jL}$}};
  \draw[quark](0.5,0)--(1.207,0.707)node[right]{{\footnotesize$e_{iL}$}};
\end{tikzpicture} & 
 \begin{tikzpicture}[]
 \draw[dashed](0,-0.5)--(0,0)node[left]{{\footnotesize$\tilde{d}_{kR}$}} -- (0,0.5);
 \draw[antiquark] (-0.977,-0.714)node[left]{{\footnotesize$\bar{u}_{jL}$}} -- (0,-0.5);
 \draw[quark] (-0.977,0.714)node[left]{{\footnotesize$\nu_{iL}$}} -- (0,0.5);
 \coordinate (A) at (0,-0.5);
  \coordinate (B) at (0,0.5);
  \coordinate (C) at (1,-0.5);
  \coordinate (D) at (1,0.5);

  \draw [antiquarku,name path=line 1] (B) -- (C)node[right]{{\footnotesize$\bar{d}_{jL}$}};

  \path[name path=line 2] (D)node[right]{{\footnotesize$e_{iL}$}} -- (A);	

  \path [name intersections={of = line 1 and line 2}];
  \coordinate (S)  at (intersection-1);

  \path[name path=circle] (S) circle(\radius);

  \path [name intersections={of = circle and line 2}];
  \coordinate (I1)  at (intersection-1);
  \coordinate (I2)  at (intersection-2);

  \draw[antiquarku] (D) -- (I1);
  \draw[-] (I2) -- (A);

  \tkzDrawArc[color=black](S,I1)(I2);

\end{tikzpicture}
\\
(a) CC $s$-channel & (b) CC $u$-channel \\
\begin{tikzpicture}[]
   \draw[quark] (-1.207,0.707)node[left]{{\footnotesize$\nu_{iL}$}} -- (-0.5,0);
  \draw[quark] (-1.207,-0.707)node[left]{{\footnotesize$d_{jL}$}} -- (-0.5,0);
  \draw[dashed](-0.5,0)--(0,0)node[above]{{\footnotesize$\tilde{d}_{kR}$}} -- (0.5,0);
  \draw[quark](0.5,0)--(1.207,-0.707)node[right]{{\footnotesize$d_{jL}$}};
  \draw[quark](0.5,0)--(1.207,0.707)node[right]{{\footnotesize$\nu_{iL}$}};
\end{tikzpicture} 
& 
 \begin{tikzpicture}[]
 \draw[dashed](0,-0.5)--(0,0)node[left]{{\footnotesize$\tilde{d}_{kR}$}} -- (0,0.5);
 \draw[antiquark] (-0.977,-0.714)node[left]{{\footnotesize$\bar{d}_{jL}$}} -- (0,-0.5);
 \draw[quark] (-0.977,0.714)node[left]{{\footnotesize$\nu_{iL}$}} -- (0,0.5);
 \coordinate (A) at (0,-0.5);
  \coordinate (B) at (0,0.5);
  \coordinate (C) at (1,-0.5);
  \coordinate (D) at (1,0.5);

  \draw [antiquarku,name path=line 1] (B) -- (C)node[right]{{\footnotesize$\bar{d}_{jL}$}};

  \path[name path=line 2] (D)node[right]{{\footnotesize$\nu_{iL}$}} -- (A);	

  \path [name intersections={of = line 1 and line 2}];
  \coordinate (S)  at (intersection-1);

  \path[name path=circle] (S) circle(\radius);

  \path [name intersections={of = circle and line 2}];
  \coordinate (I1)  at (intersection-1);
  \coordinate (I2)  at (intersection-2);

  \draw[antiquarku] (D) -- (I1);
  \draw[-] (I2) -- (A);

  \tkzDrawArc[color=black](S,I1)(I2);

\end{tikzpicture} \\
(c) NC $s$-channel & (d) NC $u$-channel \\
\begin{tikzpicture}[]
   \draw[quark] (-1.207,0.707)node[left]{{\footnotesize$\nu_{iL}$}} -- (-0.5,0);
  \draw[antiquark] (-1.207,-0.707)node[left]{{\footnotesize$\bar{d}_{kR}$}} -- (-0.5,0);
  \draw[dashed](-0.5,0)--(0,0)node[above]{{\footnotesize$\tilde{d}_{jL}^c$}} -- (0.5,0);
  \draw[antiquark](0.5,0)--(1.207,-0.707)node[right]{{\footnotesize$\bar{d}_{kR}$}};
  \draw[quark](0.5,0)--(1.207,0.707)node[right]{{\footnotesize$\nu_{iL}$}};
\end{tikzpicture} 
& 
 \begin{tikzpicture}[]
 \draw[dashed](0,-0.5)--(0,0)node[left]{{\footnotesize$\tilde{d}_{jL}^c$}} -- (0,0.5);
 \draw[quark] (-0.977,-0.714)node[left]{{\footnotesize${d}_{kR}$}} -- (0,-0.5);
 \draw[quark] (-0.977,0.714)node[left]{{\footnotesize$\nu_{iL}$}} -- (0,0.5);
 \coordinate (A) at (0,-0.5);
  \coordinate (B) at (0,0.5);
  \coordinate (C) at (1,-0.5);
  \coordinate (D) at (1,0.5);

  \draw [quarku,name path=line 1] (B) -- (C)node[right]{{\footnotesize${d}_{kR}$}};

  \path[name path=line 2] (D)node[right]{{\footnotesize$\nu_{iL}$}} -- (A);	

  \path [name intersections={of = line 1 and line 2}];
  \coordinate (S)  at (intersection-1);

  \path[name path=circle] (S) circle(\radius);

  \path [name intersections={of = circle and line 2}];
  \coordinate (I1)  at (intersection-1);
  \coordinate (I2)  at (intersection-2);

  \draw[antiquarku] (D) -- (I1);
  \draw[-] (I2) -- (A);

  \tkzDrawArc[color=black](S,I1)(I2);

\end{tikzpicture} \\
(e) NC $s$-channel & (f) NC $u$-channel
\end{tabular}
\caption{Feynman diagrams for the CC and NC contributions to the neutrino--nucleon interactions induced by the $\lambda'_{ijk}$-terms in Eq.~\eqref{laglamp}. The corresponding diagrams for the antineutrino--nucleon interactions are not shown here.}
\label{fig1}
\end{figure}

We start with the $\lambda'$-part of the RPV Lagrangian, after expanding the superpotential~\eqref{supRPV} in terms of the superfield components: 
\begin{align}
& {\cal L}_{LQD}  \ =  \   \lambda'_{ijk} \bigg[\tilde{\nu}_{iL} \bar{d}_{kR} d_{jL}  
+ \tilde{d}_{jL}  \bar{d}_{kR} \nu_{iL} 
+ \tilde{d}_{kR}^* \bar{\nu}_{iL}^c d_{jL} 
\nonumber \\
& \qquad \qquad 
- \tilde{e}_{iL} \bar{d}_{kR}  u_{jL} 
  - \tilde{u}_{jL}  \bar{d}_{kR}  e_{iL}  
- \tilde{d}_{kR}^* \bar{e}_{iL}^c u_{jL}  \bigg] +{\rm H.c.} 
\label{laglamp}
\end{align}
At the IceCube, these interactions will contribute to both charged-current (CC) and neutral-current (NC) processes mediated by either $s$-channel or $u$-channel exchange of a down-type squark, as shown in Fig.~\ref{fig1}. 

The $s$-channel processes in Figs.~\ref{fig1}(a) and (c) involve valence quarks, thus giving the dominant contributions to the (anti)neutrino--nucleon cross sections, provided the right-handed down-type squarks are produced resonantly. Similarly, the $s$-channel process in Fig.~\ref{fig1}(e) mediated by a left-handed down-type squark can also give a resonant enhancement to the (anti)neutrino--nucleon cross section. Here we have implicitly assumed that the $R$-parity conserving  (RPC) squark decays to a quark and a gluino, neutralino or chargino are suppressed~\cite{Butterworth:1992tc}, compared to the RPV decays induced by Eq.~\eqref{laglamp}.\footnote{This is the case, for instance, in the region of RPV MSSM parameter space, where the gaugino masses $M_1$, $M_2$, as well as the $\mu$-term, are larger than the squark masses, thus kinematically forbidding the two body RPC decays of squarks. The 3-body RPC decays via virtual gauginos will in general be smaller compared to the 2-body decays through RPV couplings, as considered here.} On the other hand, the contributions from the $u$-channel processes in Fig.~\ref{fig1}(b), (d) and (f) are much smaller, since they do not have a resonant enhancement, and in addition, for (b) and (d),  due to the sea quark involvement. Moreover, the RPV contributions will be sizable only for the first generation quarks, which are the predominant constituents of the nucleon, and to some extent, for the second-generation quarks. Therefore, we will ignore the contributions from the third-generation quarks. For the SM CC and NC interactions~\cite{Gandhi:1995tf, CooperSarkar:2011pa}, which must be included in the total neutrino-nucleon cross section giving rise to the IceCube events, we take into account all valence and sea quark contributions.  

The total differential cross section for the neutrino-nucleon interactions, written in terms of the Bjorken scaling variables $x=Q^2/2m_NE'_\nu$ and $y=E'_\nu/E_\nu$, is 
\begin{align}
\frac{d^2\sigma}{dxdy} \ = & \ \frac{m_N E_\nu}{16\pi} \sum_f \bigg[xf(x,Q^2)|a_f|^2 
\nonumber \\
& \qquad \qquad  \qquad 
+x\bar{f}(x,Q^2)|b_f|^2(1-y)^2\bigg] \; , \label{cross}
\end{align}
where $m_N=(m_p+m_n)/2$ is the average mass of the proton and neutron for an isoscalar nucleon, $-Q^2$ is the invariant momentum transfer between the incident neutrino and 
outgoing lepton, $E_\nu$ is the incoming neutrino energy, $E'_\nu=E_\nu-E_\ell$ is the energy loss in the laboratory frame, $E_\ell$ is the energy of the outgoing lepton, and $f(x,Q^2),\bar{f}(x,Q^2)$ are the parton distribution functions (PDFs) within the proton for $f$-quark and anti $f$-quark, respectively. For the CC processes shown in Figs.~\ref{fig1}(a) and (b), induced by an incoming neutrino of flavor $i$, the only non-trivial coefficients in Eq.~\eqref{cross} are respectively~\cite{Carena:1998gd}  
\begin{align}
a_{d_j}^{\rm CC} \ & = \ \frac{g^2}{Q^2+m_W^2} - \sum_k \frac{|\lambda'_{ijk}|^2}{xs-m^2_{\tilde{d}_{kR}}+im_{\tilde{d}_{kR}}\Gamma_{\tilde{d}_{kR}}}, \label{acc} \\
b_{\bar{u}_j}^{\rm CC} \ & = \ \frac{g^2}{Q^2+m_W^2} - \sum_k \frac{|\lambda'_{ijk}|^2}{Q^2-xs-m^2_{\tilde{d}_{kR}}}, \label{bcc}
\end{align}
where $s=2m_N E_\nu$ is the square of the center-of-mass energy, $g$ is the $SU(2)_L$ gauge coupling, and $m_W$ is the $W$-boson mass. It is obvious to see that in Eqs.~\eqref{acc} and \eqref{bcc}, the first term on the right-hand side is the SM CC contribution, whereas the second term is the RPV contribution. So for all SM CC processes involving $\bar{d}$ and $u$-type quarks, which do not have interference with the RPV processes, the coefficients in Eq.~\eqref{cross} are simply obtained by putting $\lambda'_{ijk}=0$ in Eqs.~\eqref{acc} and \eqref{bcc}. 

Similarly, for the NC processes shown in Figs.~\ref{fig1}(c)--(f), the only non-trivial coefficients are~\cite{Carena:1998gd}
\begin{align}
a_{d_j}^{\rm NC} \ & = \ \frac{g^2}{1-x_w}\frac{L_d}{Q^2+m_Z^2} - \sum_k \frac{|\lambda'_{ijk}|^2}{xs-m^2_{\tilde{d}_{kR}}+im_{\tilde{d}_{kR}}\Gamma_{\tilde{d}_{kR}}}, \label{anc} \\
b_{d_j}^{\rm NC} \ & = \  \frac{g^2}{1-x_w}\frac{R_d}{Q^2+m_Z^2} - \sum_k \frac{|\lambda'_{ijk}|^2}{Q^2-xs-m^2_{\tilde{d}_{kL}}}, \label{bnc}
\end{align}
where $L_d=-(1/2)+(1/3)x_w$ and $R_d=(1/3)x_w$ are the chiral couplings, $x_w\equiv \sin^2\theta_w$ is the weak mixing angle parameter and $m_Z$ is the $Z$-boson mass. For all SM NC processes involving $u$-type quarks, which do not have interference with the RPV processes, the coefficients in Eq.~\eqref{cross} are simply obtained by putting $\lambda'_{ijk}=0$ in Eqs.~\eqref{anc} and \eqref{bnc} and replacing $L_d\to L_u= (1/2)-(2/3)x_w$, $R_d\to R_u=-(2/3)x_w$.
For neutrino-antiquark interactions, the coefficients for the NC processes can be obtained simply by crossing symmetry, i.e.~$a_f\leftrightarrow b_f,\ xs\leftrightarrow Q^2-xs$. Similarly, for antineutrino-nucleon interactions, we can just replace the PDFs $f \leftrightarrow \bar{f}$ in Eq.~\eqref{cross}. 

Note the Breit-Wigner resonance form of Eqs.~\eqref{acc} and \eqref{anc}, which is regulated by the right-handed down-type squark width 
\begin{align}
\Gamma_{\tilde{d}_{kR}} \ \simeq \ \frac{m_{\tilde{d}_{kR}}}{8\pi}\sum_{ij} |\lambda'_{ijk}|^2,
\label{decaydR}
\end{align}
 assuming that the only dominant decay modes are $\tilde{d}_{kR}\to \nu_{iL}d_{jL}$ (NC) and $\tilde{d}_{kR} \to e_{iL}u_{jL}$ (CC), and the masses of the final state fermions in these 2-body decays are negligible compared to the parent squark mass. For the left-handed down-type squark, $\Gamma_{\tilde{d}_{kL}}=\Gamma_{\tilde{d}_{kR}}/2$, since $\tilde{d}_{kL}\to \nu_{iL}d_{jR}$ is the only available decay mode. The resonance condition is satisfied for the incoming energy $E_\nu=m^2_{\tilde d_{kR}}/2m_Nx$, but due to the spread in the initial quark momentum fraction $x\in [0,1]$, the resonance peak will be broadened and shifted above the threshold energy $E_\nu^{\rm th}=m^2_{\tilde d_{kR}}/2m_N$ (see Fig.~\ref{fig2}). Nevertheless, one can immediately infer that for $m_{\tilde d_{kR}}\in$ [100  GeV, 2 TeV], $E_\nu^{\rm th}$ is in the multi TeV--PeV range, and hence, can be probed by the available IceCube HESE data.     

\section{Event Rate at IceCube} \label{sec:3}
\begin{figure}[t]
\centering
\includegraphics[width=10cm]{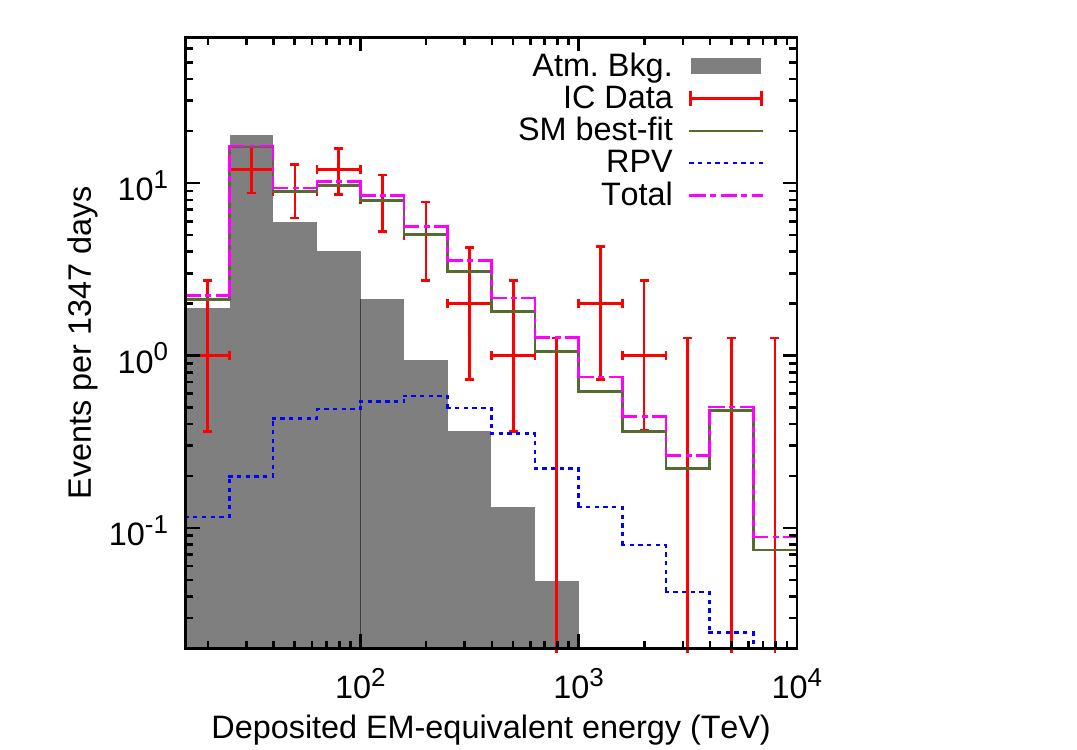}
\caption{Event distribution at the IceCube without and with an RPV contribution, and comparison with the 4-year HESE data. The shaded region shows the expected background from  atmospheric muons and muon neutrinos. For the RPV case, we have taken $m_{\tilde{d}_L}=m_{\tilde{d}_R}=400$ GeV, $|\lambda'_{11k}|=0.4$ and all other $|\lambda'_{ijk}|=0$. }
\label{fig2}
\end{figure}

The expected number of HESE events in a given deposited energy bin at IceCube due to the modified cross section~\eqref{cross} can be estimated as~\cite{Chen:2013dza}
\begin{align}
N_{\rm bin} \ = \ TN_A\int_{E^{\rm bin}_{\rm min}}^{E^{\rm bin}_{\rm max}}dE_{\rm dep} \int_0^1 dy \: V_{\rm eff} \:  \Phi \: \Omega \: \frac{d\sigma}{dy} \, ,
\label{event}
\end{align}
where $T$ is the exposure time, $N_A$ is the Avogadro number, $E_{\rm dep}(E_\nu)$ is the electromagnetic (EM)-equivalent deposited energy for a 
given incoming neutrino energy $E_\nu$ in the laboratory frame,  $V_{\rm eff}(E_\nu)$ is the effective target volume of the detector, $\Phi(E_\nu)$ is the incident neutrino flux, $\Omega(E_\nu)$ is the effective solid angle of coverage, and we have integrated the differential cross section in Eq.~\eqref{cross} over $x\in [0,1]$, including both neutrino and antineutrino initial states with all flavors;  for details, see Refs.~\cite{Aartsen:2013bka, Aartsen:2015zva, Kistler:2013my, Laha:2013lka, Chen:2013dza, Vincent:2016nut}. As an illustration, we show in Fig.~\ref{fig2} the predicted number of events with and without RPV interactions for 1347 days of exposure at IceCube and compare them to the corresponding 4-year HESE data in each of the 14 deposited energy bins. Here we have assumed the IceCube best-fit astrophysical power-law flux $E^2\Phi(E) = 2.2\times 10^{-8}(E/100~{\rm TeV})^{-0.58}~{\rm GeV}\: {\rm cm}^{-2}{\rm s}^{-1}{\rm sr}^{-1}$~\cite{Aartsen:2015zva} with a standard $(1:1:1)$ flavor composition ratio on Earth, and have used the NNPDF2.3 leading order PDF sets~\cite{Ball:2012cx} for the cross section calculations.\footnote{The PDF uncertainties on the total cross-section are at most at 5\% level~\cite{Chen:2013dza} for the energy range of interest. Therefore, we only consider the central values of the cross sections. The flux uncertainties, on the other hand, are currently at the level of 15\%~\cite{Aartsen:2015zva}.} To illustrate the RPV contribution, we have  considered 
 $m_{\tilde{d}_L}=m_{\tilde{d}_R}=400$ GeV, $|\lambda'_{11k}|=0.4$, and 
all other $|\lambda'_{ijk}|=0$ as our benchmark point. 

In Fig.~\ref{fig2}, the IceCube data points, as well as the background due to atmospheric 
neutrinos and muons, are taken from Ref.~\cite{Aartsen:2015zva}.\footnote{We have not considered here the latest through-going 
track signal with $E_{\rm dep}=2.6\pm 0.3$ PeV~\cite{track}, since it was not included in the analysis of Ref.~\cite{Aartsen:2015zva}.}  Note that 
atmospheric $\nu_e$ events will also get modified due to 
nonzero RPV couplings $|\lambda'_{11k}|$. However, since the atmospheric 
background is dominated by the $\nu_\mu$-induced events and the event rate 
for atmospheric $\nu_e$ is much smaller~\cite{Aartsen:2015xup},  we can 
safely ignore the $|\lambda'_{11k}|$ effects on the background and just 
assume it to be basically the same as in the SM case. Also one might wonder 
whether the source flavor composition and flux of the neutrinos could be 
modified due to the new RPV interactions. However, this effect is expected 
to be small for the values of squark masses and RPV couplings considered here, 
since the SM weak interactions with strength 
$G_F$ (the Fermi coupling constant) will be dominant over the RPV 
interactions with relative strength of $|\lambda'_{ijk}|^2/m_{\tilde d_k}^2$. 
The RPV effect at the IceCube detector could get enhanced {\it only} due to the resonant 
production of squarks in a conducive range of the incoming neutrino energy. Thus, adding the flux uncertainty in our analysis will equally affect the events due to both SM and RPV interactions, without changing the relative enhancement of the events in presence of the RPV interactions with respect to the SM prediction. This justifies our use of the IceCube best-fit value for the flux.   

From Fig.~\ref{fig2}, one can see the small enhancement in the total number of events over the SM prediction in presence of RPV SUSY. A larger $|\lambda'_{11k}|$ will result in a more pronounced excess in some of the energy bins. Our benchmark point was partly motivated by the fact that there seems to be a small excess in the data around 100 TeV, which could in principle be explained by our RPV scenario, if it becomes statistically significant. There seems to be another excess just above 1 PeV, which is however difficult to explain in this scenario, since this would require a squark mass above TeV, for which the production cross section is already small. Moreover, since the neutrino flux has a strong power-law dependence of $E^{-2.58}$, the resulting number of events in the higher-energy bins will be further suppressed, thus requiring a non-perturbative value of $|\lambda'_{11k}|$ to fit any PeV-excess.

\section{Correlation with $0\nu\beta\beta$}\label{sec:cor}
For $k=1$, a larger $|\lambda'_{11k}|$-coupling will also enhance the rate of $0\nu\beta\beta$ in nuclei, thus giving a smaller lifetime and a {\it negative} correlation with the event rate at IceCube. Including only the $|\lambda'_{111}|$-diagrams and ignoring all other RPV contributions~\cite{Rodejohann:2011mu} for simplicity, we can write down the expression for the $0\nu\beta\beta$ half-life as 
\begin{align}
\frac{1}{T_{1/2}^{0\nu}} \ = \ G_{01}\bigg|\frac{m_{\beta\beta}}{m_e}M_\nu+e^{i\phi}M_{\lambda'_{111}} \bigg|^2 ,
\label{half}
\end{align}
where $G_{01} = 5.77\times 10^{-15}~{\rm yr}^{-1}$ is the phase space factor for $^{76}$Ge~\cite{Kotila:2012zza} (which is taken here as our benchmark nucleus), $m_e$ is the electron mass, $m_{\beta\beta}$ is the effective mass corresponding to the light neutrino contribution with the nuclear matrix element (NME) $M_\nu$ and $\phi$ is a relative phase between the light neutrino and RPV contributions. The explicit form of the NME for the RPV contribution is~\cite{Hirsch:1995zi, Faessler:1998qv}
\begin{align}
M_{\lambda'_{111}} \  =  &\  (\eta_{\tilde{g}}+\eta_\chi)M_{\tilde{g}}^{2N}+(\eta_{\chi \tilde{e}}+\eta'_{\tilde{g}}+\eta_{\chi \tilde{f}})M_{\tilde{f}}^{2N} \nonumber \\
& 
+\frac{3}{8}\bigg[(\eta_{\tilde{g}}+\eta_\chi)\nonumber \\ & 
+\frac{5}{3}(\eta_{\tilde{g}}+\eta_\chi+\eta_{\chi \tilde{e}}+\eta'_{\tilde{g}}+\eta_{\chi \tilde{f}})\bigg] \nonumber \\
& \qquad \qquad \times 
\bigg(\frac{4}{3}M^{1\pi}+M^{2\pi}\bigg) ,
\label{nme}
\end{align}  
where the amplitudes of the different RPV contributions are given by~\cite{Allanach:2009xx} 
\begin{align}
\eta_{\tilde{g}} \ & = \ \frac{\pi \alpha_s}{6}\frac{\lambda'^2_{111}}{G_F^2}\frac{m_p}{m_{\tilde{g}}}\bigg[\frac{1}{m^4_{\tilde{u}_L}}+ \frac{1}{m^4_{\tilde{d}_R}} -\frac{1}{2m^2_{\tilde{u}_L}m^2_{\tilde{d}_R}}\bigg], \nonumber \\ 
\eta_\chi \ & = \ \frac{\pi \alpha_2}{2}\frac{\lambda'^2_{111}}{G_F^2}\sum_{i=1}^4 \frac{m_p}{m_{\chi_i}}\bigg[\frac{\epsilon^2_{L_i}(u)}{m^4_{\tilde{u}_L}}+\frac{\epsilon^2_{R_i}(d)}{m^4_{\tilde{d}_R}}-\frac{\epsilon_{L_i}(u)\epsilon_{R_i}(d)}{m^2_{\tilde{u}_L}m^2_{\tilde{d}_R}} \bigg], \nonumber \\ 
\eta_{\chi \tilde{e}} \ & = \ 2\pi \alpha_2\frac{\lambda'^2_{111}}{G_F^2}\sum_{i=1}^4 \frac{m_p}{m_{\chi_i}}\frac{\epsilon^2_{L_i}(e)}{m^4_{\tilde{e}_L}}, \nonumber   \\ 
\eta'_{\tilde{g}} \ & = \ \frac{2\pi\alpha_s}{3}\frac{\lambda'^2_{111}}{G_F^2}\frac{m_p}{m_{\tilde{g}}}\frac{1}{m^2_{\tilde{u}_L}m^2_{\tilde{d}_R}} ,\nonumber \\
\eta_{\chi \tilde{f}} \ & = \ \pi \alpha_2 \frac{\lambda'^2_{111}}{G_F^2}\sum_{i=1}^4 \frac{m_p}{m_{\chi_i}}\bigg[\frac{\epsilon_{L_i}(u)\epsilon_{R_i}(d)}{m^2_{\tilde{u}_L}m^2_{\tilde{d}_R}}-\frac{\epsilon_{L_i}(u)\epsilon_{L_i}(e)}{m^2_{\tilde{u}_L}m^2_{\tilde{e}_L}} \nonumber \\
& \qquad 
-\frac{\epsilon_{L_i}(e)\epsilon_{R_i}(d)}{m^2_{\tilde{e}_L}m^2_{\tilde{d}_R}} \bigg], 
\label{ampl}
\end{align} 
and $\epsilon$'s denote rotations between the mass and gauge eigenbasis in the gaugino-fermion-sfermion vertices. For $^{76}$Ge, the NMEs for the light neutrino, 2-nucleon, 1-pion and 2-pion exchange modes are respectively given by~\cite{Pantis:1996py, Hirsch:1995zi, Faessler:1998qv}
\begin{align}
& M_{\nu} \ = \ 2.8, \quad M_{\tilde{g}}^{2N} \ = \ 283,  \quad 
M_{\tilde{f}}^{2N} \ = \ 13.2, \nonumber \\
& M^{1\pi} \ = \ -18.2, \quad 
M^{2\pi} \ = \ -601.  
\end{align} 
\begin{figure}[t!]
\centering
\includegraphics[width=8cm]{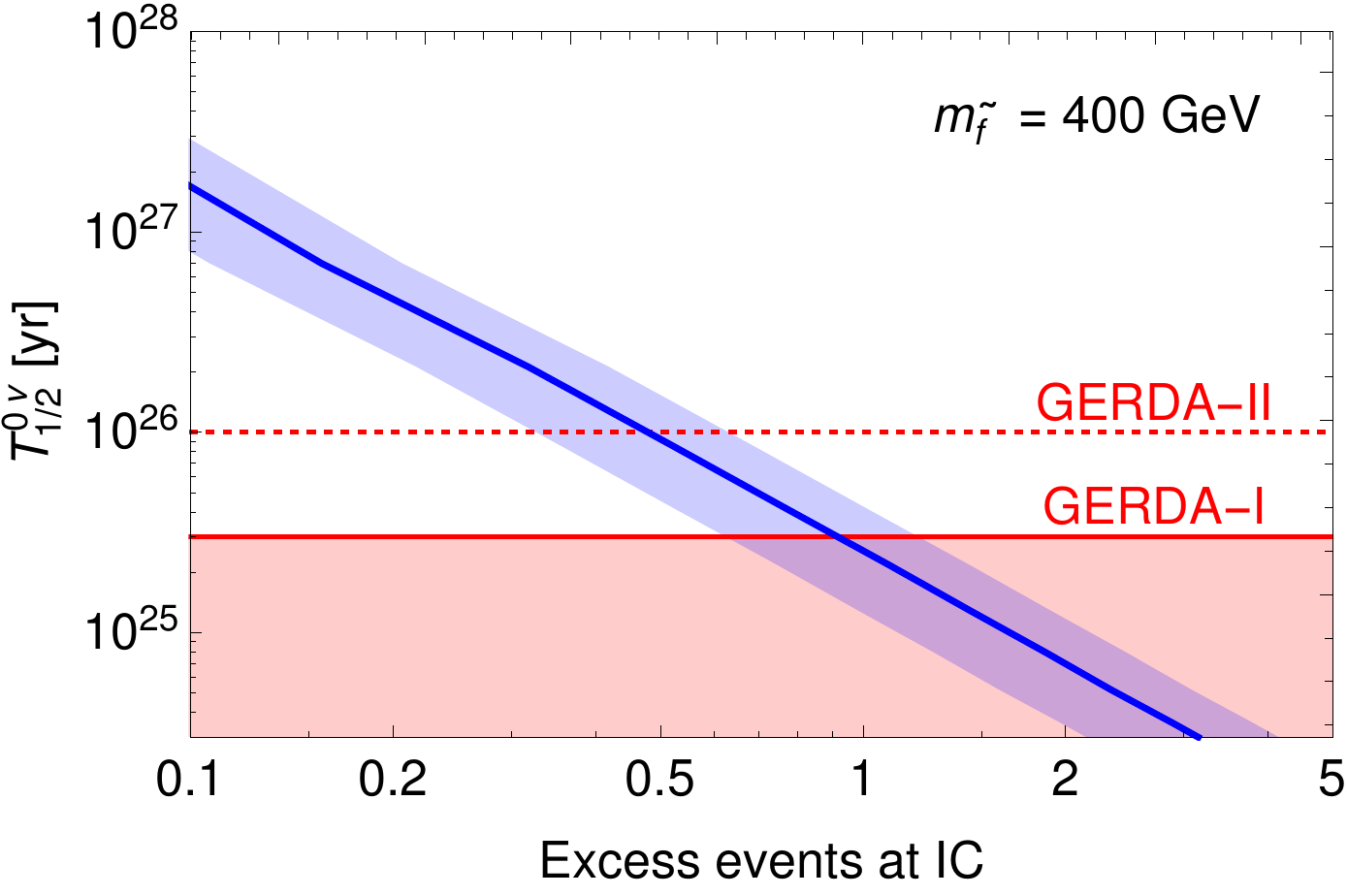}
\caption{Correlation between the $0\nu\beta\beta$ half-life of $^{76}$Ge and total number of excess events (over the SM prediction) at the IceCube in our RPV scenario. Here we have chosen all the  relevant  sfermion masses appearing in Eqs.~\eqref{ampl} to be 400 GeV, while all the gaugino masses in Eqs.~\eqref{ampl} are assumed to be much larger (around 10 PeV) to avoid the stringent $0\nu\beta\beta$ limits in part of the parameter space shown here. The (blue) solid curve is obtained by varying $\lambda'_{111}$, while keeping all other RPV couplings zero. The blue band corresponds to the $1\sigma$ uncertainty in the neutrino flux.}
\label{fig3}
\end{figure}
Using Eqs.~\eqref{event} and~\eqref{half}, we show in Fig.~\ref{fig3} the correlation between the $0\nu\beta\beta$ lifetime and the {\it total} number of excess events over the SM expectations (summed over all the bins shown in Fig.~\ref{fig2}), as predicted with 4-year exposure at the IceCube in our RPV scenario. For illustration, we have chosen here $m_{\beta\beta}=5$ meV so that the RPV contribution is the dominant one for the experimentally accessible range and the relative phase $\phi$ in Eq.~\eqref{half} does not play any role. For larger values of $m_{\beta\beta}$, one could have either a constructive or destructive interference between the light neutrino and RPV contributions in Eq.~\eqref{half}, depending on the phase $\phi$. For the SUSY spectrum, we have fixed the squark and slepton masses entering into Eqs.~\eqref{ampl} at a common value of 400 GeV, whereas the gluino and neutralinos are assumed to be much heavier around 10 PeV, in order to avoid the stringent limits on $0\nu\beta\beta$ half-life in at least part of the parameter space shown in Fig.~\ref{fig3}, which could still yield an observable excess at the IceCube. The blue, solid curve is obtained by varying $\lambda'_{111}$ (from 0.1 to 2), whereas the band around this curve shows the $1\sigma$ uncertainty in the best-fit neutrino flux~\cite{Aartsen:2015zva}.\footnote{The band will be broader  if we also include the NME uncertainties (which are expected to be 50\% or even higher~\cite{Meroni:2012qf}), but due to the interplay of different NMEs (of both signs) in Eq.~\eqref{nme}, calculating the total NME uncertainty is highly non-trivial, and hence, not pursued here.} We find that for our chosen benchmark values of the squark and gaugino masses, $|\lambda'_{111}|\gtrsim 0.7$ (marked by the intersection of the blue and red solid lines in Fig.~\ref{fig3}) is excluded from the combined lower limit of $T_{1/2}^{0\nu}>3\times 10^{25}$ yr from GERDA phase-I+Heidelberg-Moscow+IGEX~\cite{Agostini:2013mzu}, and this could be improved up to $|\lambda'_{111}|\gtrsim 0.5$ (marked by the intersection of the blue solid and red dashed lines in Fig.~\ref{fig3}) with the projected sensitivity of GERDA phase-II~\cite{Andrea:2016lfr}. Thus, Fig.~\ref{fig3} implies that we cannot expect a large $\lambda'_{111}$-contribution to the current or future IceCube HESE data due to the $0\nu\beta\beta$ constraints. Similar conclusions can be derived for other $\lambda'_{ijk}$-contributions by considering the corresponding limits from other low-energy processes, as demonstrated in the following section.  

In principle, one can also consider the RPV-assisted long range contributions to $0\nu\beta\beta$, mediated by a light neutrino and a squark, which are independent of the gaugino mass. This contribution is mostly relevant for the coupling product $\lambda'_{113}\lambda'_{131}$, which depends on the left-right sbottom mixing matrix. However, it is strongly constrained by $B$-physics and light neutrino mass observables~\cite{Allanach:2009xx}, and therefore, can not give rise to a significant effect at IceCube.

\section{Upper limit on $|\lambda'_{ijk}|$} \label{sec:4}
Since no statistically significant excess over the SM prediction is seen in the current IceCube data (cf. Fig.~\ref{fig2}), we use this information to put an upper bound on the $|\lambda'_{ijk}|$ couplings. To this effect, we perform a binned likelihood analysis~\cite{Aartsen:2014muf} with the Poisson likelihood function
\begin{align}
L \ = \ \prod_{{\rm bins}~i} \frac{e^{-\lambda_i}\lambda_i^{n_i}}{n_i!},
\end{align}
where the observed count $n_i$ in each bin $i$ is compared to the theory prediction 
$\lambda_i$, including the RPV contribution induced by $\lambda'_{ijk}$. We then construct a test statistic 
\begin{align}
-2\Delta \ln L \ = \ -2(\ln L-\ln L_{\rm max}),
\end{align}
from which a $1\sigma$ ($2\sigma$)  upper limit on $|\lambda'_{ijk}|$ corresponding to the value of $-2\Delta \ln L=1~(2.71)$ can be derived. Here $L_{\rm max}$ represents the likelihood value obtained with $\lambda'_{ijk}=0$, i.e. including only the SM contribution in the analysis. 

\begin{figure*}[t]
\centering
\begin{tabular}{cc}
\includegraphics[width=7cm]{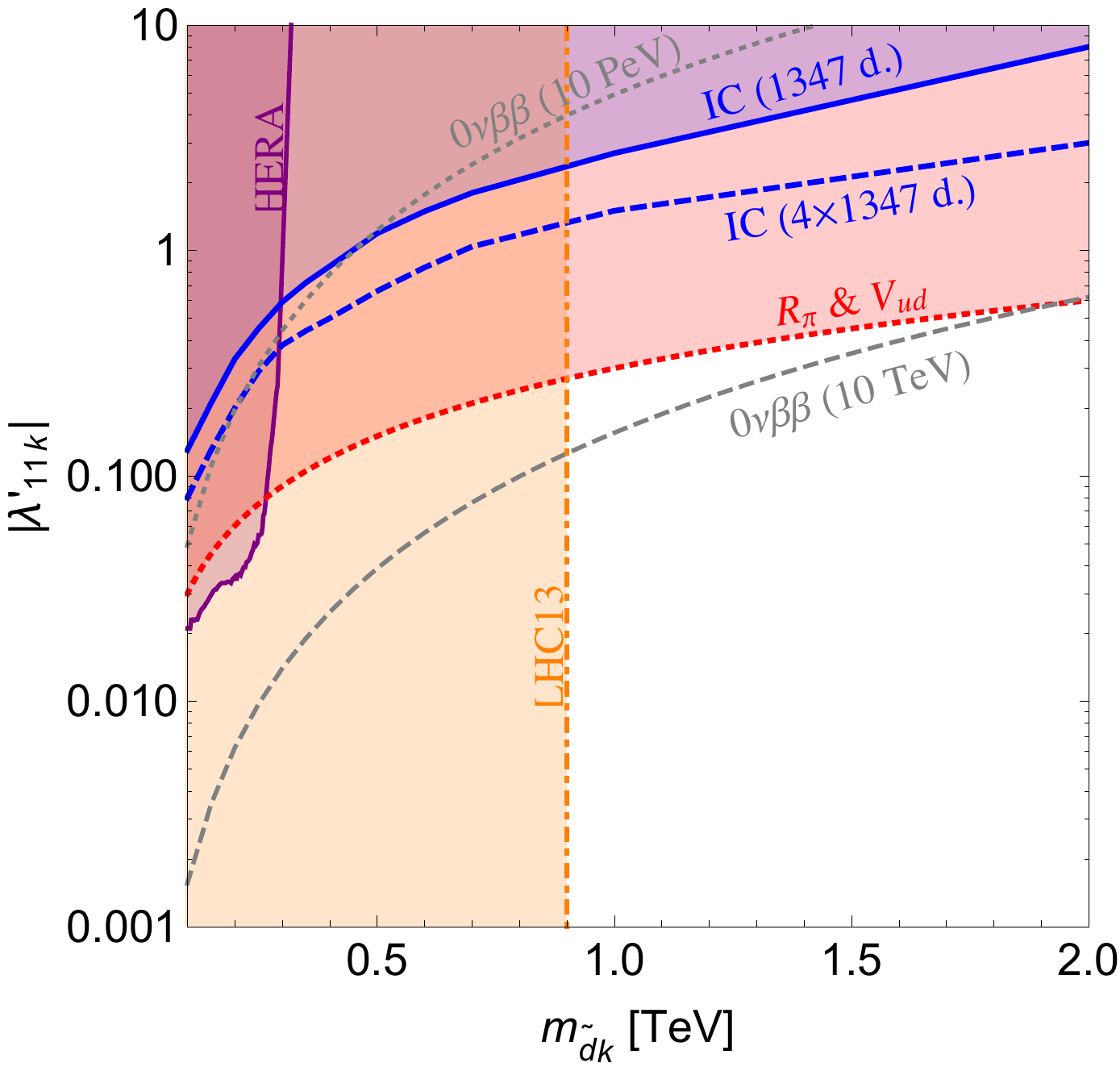} & 
\includegraphics[width=6.5cm]{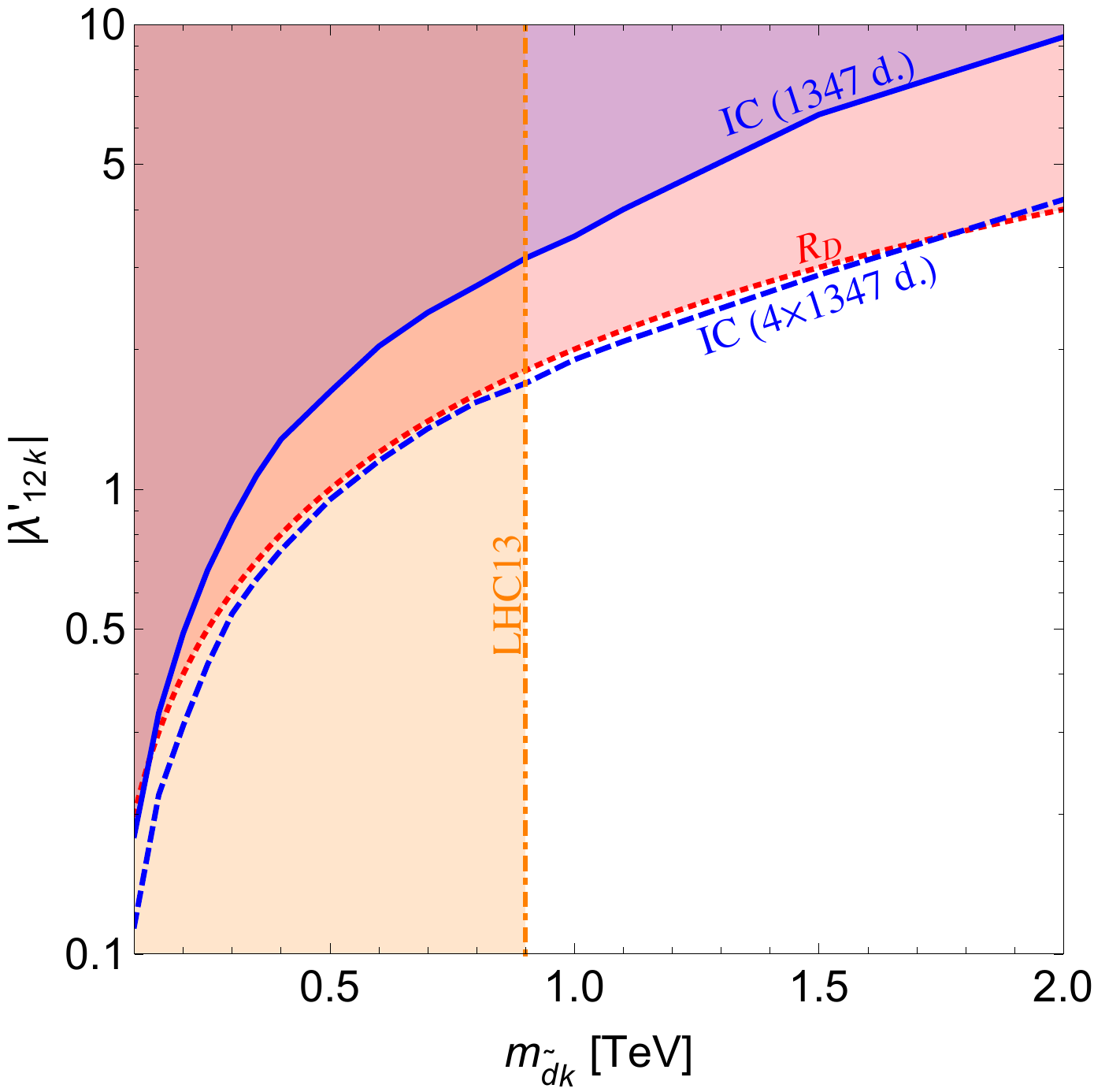} \\
(a) $|\lambda'_{11k}|$ & (b) $|\lambda'_{12k}|$ 
\end{tabular}
\caption{$1\sigma$ upper limits (blue, solid) on the RPV couplings $|\lambda'_{1jk}|$ (with $j=1,2$) from the 4-year IceCube HESE data and projected limits (blue, dashed) obtained by scaling the exposure time by a factor of 4. For comparison, we also show the $2\sigma$ indirect limits from lepton universality in meson decays (red, dotted) and the 95\% CL direct limits from a scalar leptoquark search at the 13 TeV LHC (orange, dot-dashed). In (a), we additionally show the 95\% CL direct search limit (magenta, solid) from $e^-p$ collisions at HERA, as well as the 90\% CL $0\nu\beta\beta$ limits for two benchmark values of 10 TeV (gray, dashed) and 10 PeV (gray, dotted) for the gaugino masses appearing in Eqs.~\eqref{ampl}, while keeping all the relevant sfermion masses fixed at $m_{\tilde{d}_k}$.}\label{fig4}
\end{figure*}

Our results for the conservative $1\sigma$ limits are shown in Fig.~\ref{fig4} (blue solid curves) for $|\lambda'_{11k}|$ and $|\lambda'_{12k}|$,~i.e. for electron-type neutrinos interacting with the 1st and 2nd generation quarks, respectively. As expected, the limits for the 1st generation are stronger, since the corresponding cross sections are larger due to the large valence-quark content of the nucleon at the high values of Bjorken-$x$ required to resonantly produce squarks of significant mass. 

There exist stringent limits on $|\lambda'_{11k}|$ from direct searches in $e^\pm p$ collisions at HERA with $\sqrt s=319$ GeV~\cite{Aaron:2010ez, South:2016cmx}, as shown by the magenta-shaded region in Fig.~\ref{fig4}(a). Squark masses below 100 GeV or so are disfavored from direct searches for RPV SUSY at LEP~\cite{Abbiendi:2001aj, Heister:2002jc, Abbiendi:2003rn}, Tevatron~\cite{Abbott:1999nh, Abe:1998gu} and LHC~\cite{atlas-susy, cms-susy, Redelbach:2015meu}, and therefore, are not considered here.\footnote{In the absence of the possibility of a resonant production (as e.g. in the sneutrino case) in $e^+e^-$ and hadron-hadron collisions, it is difficult to cast most of the collider limits onto the $m_{\tilde{d}}-|\lambda'_{ijk}|$ plane in a model-independent manner, and therefore, we do not attempt to show them in Fig.~\ref{fig4}.} In addition, the recent search for scalar leptoquarks at the 13 TeV LHC with $3.2~{\rm fb}^{-1}$ data~\cite{Aaboud:2016qeg} is relevant for our RPV scenario, since $\lambda'_{ijk}$-couplings also give rise to the same $e_i e_ijj$ final states via pair-production of down-type squarks, followed by $\tilde{d}_{kR}\to e_{iL}u_{jL}$ which has a branching ratio of 0.5. The corresponding 95\% CL ATLAS limit of 900 GeV on the first-generation scalar leptoquark mass can be directly translated into a lower bound on the down-type squark mass, as shown by the vertical dot-dashed line in Fig.~\ref{fig4}.  There also exist indirect constraints on $|\lambda'_{11k}|$ from lepton universality in pion decay, measured by the ratio $R_\pi=\frac{{\rm BR}(\pi^-\to e^-\bar{\nu}_e)}{{\rm BR}(\pi^-\to \mu^-\bar{\nu}_\mu)}$, unitarity of the CKM element $V_{ud}$ and atomic parity violation~\cite{Kao:2009fg}, the most stringent of which is shown in Fig.~\ref{fig4}(a) by the red dotted curve. Other low-energy constraints, such as neutrino mass~\cite{Bhattacharyya:1999tv}, electric dipole moment~\cite{Yamanaka:2014nba} and flavor-changing $B$-decays~\cite{Ghosh:2001mr, Deshpande:2004xc, Dreiner:2013jta}, always involve the product of two independent RPV couplings, and hence, are not applicable in our case. Moreover, for $k=1$, we have an additional constraint from $0\nu\beta\beta$, as discussed in the previous section, which however depends on the masses of other SUSY particles, unlike the other limits discussed above, which are independent of the rest of the SUSY spectrum, as long as the 2-body RPV decay modes of the squark are dominant. Just for the sake of comparison, we show the $0\nu\beta\beta$ limits in Fig.~\ref{fig4}(a) for two benchmark points with $m_{\tilde{g}}=m_{\chi_i}=10$ TeV (gray, dashed) and 10 PeV (gray, dotted), while keeping all sfermion masses equal to $m_{\tilde{d}}$. In the former case, the $0\nu\beta\beta$ limit is the most stringent one, whereas for either heavier gaugino masses or $k\neq 1$, the pion decay constraint is stronger than the limit obtained from IceCube in the entire mass range considered. Similarly, as shown in Fig.~\ref{fig4}(b), for $|\lambda'_{12k}|$, the indirect constraints from lepton universality in neutral and charged $D$-meson decays, measured by the ratio $R_D=\frac{{\rm BR}(D\to Ke\nu_e)}{{\rm BR}(D\to K\mu\nu_\mu)}$,  
are stronger than the IceCube limit derived here. This rules out the possibility of any $|\lambda'_{1jk}|$-induced observable excess in the 4-year IceCube data.

However, one should note that these are the first-ever IceCube constraints on RPV couplings, and at present, are mostly limited by statistics, which is  expected to improve 
significantly with more exposure time. To illustrate this point, we just 
scale the current 4-year dataset by a factor of 4 (roughly corresponding to 
15 years of actual data taking) in all the bins analyzed here and derive 
projected limits on $|\lambda'_{1jk}|$ (with $j=1,2$), following the same 
likelihood procedure described above. This conservative estimate of the future limits is shown in Fig.~\ref{fig4} by 
the blue dashed curves. We find that the limit on $|\lambda'_{1jk}|$ 
can be improved roughly by up to a factor of 3 with 15-yr IceCube data, and it might even surpass the current best limit in the sub-TeV squark mass range for $j=2$, although the indirect limit from lepton universality could improve by an order of magnitude in a future super-tau-charm factory~\cite{Eidelman:2015wja}. In practice, however, we may not have to 
wait for 15 years, since a number of unforeseen factors could improve the 
conservative projected IceCube limits shown here, e.g. the future data in all the bins may not scale proportionately to the current data and may turn out to be in better agreement
 with the SM prediction. Similarly, other large-volume detectors like KM3Net and 
IceCube-Gen2 might go online at some point,  thus significantly increasing the total statistics. 

We also note that a similar analysis could be performed for incident neutrinos of muon and tau flavors at IceCube, though for muon neutrinos, one has to carefully reassess the atmospheric background including the RPV effects. However, we expect the corresponding limits on $|\lambda'_{ijk}|$ (with $i=2,3$ and $j=1,2$) to be weaker than the limits on $|\lambda'_{1jk}|$ shown in Fig.~\ref{fig4} simply due to the fact that the effective fiducial volume at the IceCube is the largest for $\nu_e$~\cite{Aartsen:2013bka}. Nevertheless, with more statistics, one could in principle consider the $|\lambda'_{ijk}|$ couplings for all neutrino flavors. Also, one could improve the analysis presented here by taking into account the showers and tracks individually. Since the $|\lambda'_{1jk}|$ couplings preferentially enhance {\it only} one type of events, viz. showers for $i=1,3$ and tracks for $i=2$, a binned track-to-shower ratio analysis is expected to improve the limits on the corresponding $|\lambda'_{ijk}|$. In fact, by examining the track-to-shower ratio in future data, one might be able to distinguish between different new physics contributions to the IceCube events, provided the source flavor composition of the neutrinos is known more accurately. 

Before concluding, we would like to make a comment on the $\lambda$-couplings in Eq.~\eqref{supRPV}, which leads to the $LLE$-type RPV Lagrangian
\begin{align}
{\cal L}_{LLE} \ = & \ \frac{1}{2}\lambda_{ijk}\bigg[ \tilde{\nu}_{iL}\bar{d}_{kR}d_{jL}+\tilde{e}_{iL}\bar{e}_{kR}\nu_{jL}+\tilde{e}^*_{kR}\bar{\nu}^c_{iL}e_{jL} \nonumber \\
& \qquad \qquad 
-(i\leftrightarrow j)\bigg]+{\rm H.c.}
\end{align}
This will give rise to a selectron resonance from (anti)neutrino--electron interactions at IceCube. However, the corresponding threshold energy $E_\nu^{\rm th}=m^2_{\tilde{e_k}}/2m_e$ is beyond 10 PeV for selectron masses above 100 GeV or so. Since smaller selectron masses are excluded from the LEP data~\cite{Abbiendi:2003rn, Abdallah:2003xc}, we cannot probe the $\lambda_{ijk}$-couplings with the current IceCube data. Nevertheless, if future data reports any events beyond 10 PeV, the $LLE$-type RPV scenario could in principle provide a viable explanation, given the fact that it would be  difficult to explain those events within the SM and with an unbroken power-law flux, without having a significantly larger number of events in all the preceding lower-energy bins.

\section{Conclusion} \label{sec:5}
RPV SUSY is a well-motivated candidate for TeV-scale new physics beyond the SM, while being consistent with the null results at the LHC so far. Therefore, it is important to test this hypothesis at different energy scales available to us. Using the 4-year IceCube HESE data in the multi TeV-PeV range, we have derived the first IceCube upper limits on the RPV couplings $|\lambda'_{ijk}|$ as a function of the mass of the resonantly-produced down-type squarks (see Fig.~\ref{fig4}). Although weaker than the existing limits from low-energy processes, our limits are expected to be significantly improved with more statistics in future, thereby complementing the RPV SUSY searches at the energy and intensity frontiers. 

\section*{Acknowledgments}
B.D. is grateful to Chien-Yi Chen and Amarjit Soni for many useful 
discussions on IceCube. The work of B.D. is supported by the DFG grant 
RO 2516/5-1 and W.R. is supported by the 
DFG in the Heisenberg programme with grant RO 2516/6-1.  B.D. also acknowledges partial support from the 
TUM University Foundation Fellowship, the DFG cluster of excellence 
``Origin and Structure of the Universe", and the Mainz Institute for Theoretical Physics during various stages of this work. 
D.K.G. would like to thank Texas A \& M University for the hospitality, where
the final stage of this work was done.

\end{document}